\documentclass[10pt,a4paper]{article}

\def\tr{\;{\rm tr}\;}

\newcommand{\beq}{\begin{equation}}
\newcommand{\eeq}{\end{equation}}

\newcommand{\beqs}{\begin{eqnarray}}
\newcommand{\eeqs}{\end{eqnarray}}

\newcommand{\half}{\frac{1}{2}}

\newcommand{\ov}[1]{\frac{1}{#1}}

\newcommand{\Ntr}{{{\rm tr} \over N}}

\newcommand{\fl}{\noindent}

\newtheorem{sutra}{}

\newtheorem{bhashya}{}[sutra]

\begin{document}
\input{epsf}

\title{\bf \large Variational ansatz for gaussian + Yang-Mills
two matrix model compared with Monte-Carlo simulations
in 't Hooft limit}

\author{Govind S.  Krishnaswami \\
    Department of Physics and Astronomy,\\
    University of Rochester, Rochester, New York 14627 \\
   }

\maketitle

\begin{abstract}

In recent work, we have developed a variational principle for
large $N$ multi-matrix models based on the extremization of
non-commutative entropy. Here, we test the simplest variational
ansatz for our entropic variational principle with Monte-Carlo
measurements. In particular, we study the two matrix model with
action $\tr [{m^2 \over 2} (A_1^2 + A_2^2) -\ov{4} [A_1,A_2]^2]$
which has not been exactly solved. We estimate the expectation
values of traces of products of matrices and also those of traces
of products of exponentials of matrices (Wilson loop operators).
These are compared with a Monte-Carlo simulation. We find that the
simplest wignerian variational ansatz provides a remarkably good
estimate for observables when $m^2$ is of order unity or more. For
small values of $m^2$ the wignerian ansatz is not a good
approximation: the measured correlations grow without bound,
reflecting the non-convergence of matrix integrals defining the
pure commutator squared action.
Comparison of this ansatz with the exact solution of a two matrix model
studied by Mehta is also summarized. Here the wignerian ansatz is
a good approximation both for strong and weak coupling.

\end{abstract}

KEYWORDS: multi matrix models, Yang-Mills integrals, Yang-Mills
theory, M(atrix) theory, large N limit, entropy, variational
principle, Monte-Carlo integration.

\pagebreak

\section{Introduction}

Matrix models are of long standing interest in several branches of
physics and mathematics.

Early work of Wigner, Dyson and others introduced random matrices
in the study of statistical properties of highly excited energy
levels of nuclei \cite{mehta-book}.

In the early 1970s the work of 't Hooft
\cite{thooft-book,witten-one-over-n} on the large $N$ limit of QCD
($N$ is the number of colors, gluon fields are $N \times N$
matrices), made the study of large $N$ matrix field theories a
central theme in understanding the non-perturbative dynamics of
non-abelian gauge theories. This also gave the first indication
that it is in the large $N$ limit that a gauge theory may have the
long sought dual description as a string theory.

Important progress was made in the late 1970s and early 1980s
stemming from Migdal and Makeenko's work on the factorized loop
equations of large $N$ gauge theories \cite{migdal-loop}. The work
of Sakita and Jevicki \cite{jevicki-sakita} on the collective
field formalism of large $N$ field theories and that of Cvitanovic
and collaborators \cite{cvitanovic} also bears mention from this
period. Eguchi and Kawai's \cite{eguchi-kawai} proposal on
reducing a matrix field theory to a matrix model with a finite
number of degrees of freedom but in the large $N$ limit has been a
recurring theme ever since.

Important breakthroughs in the study of random surfaces, two
dimensional string theory and two dimensional gravity coupled to
matter were made in the late 80s and early 1990s (see
\cite{ginsparg-moore} for a review) The planar Feynman graph
expansion of large $N$ matrix models was used as a way of
descretizing a two dimensional surface. Models with one or a
finite number of matrices and the $c=1$ quantum mechanics of a
single matrix were of importance in these developments. In
addition to the large $N$ limit, the double scaling limit was
developed to study the surfaces obtained in the continuum limit.

From the early 1990s onwards, the work of mathematicians including
Voiculescu and collaborators on von Neumann algebras lead to the
development of the field of non-commutative probability theory
\cite{voiculescu}. Large $N$ matrix models are natural examples of
non-commutative probability theories. Random matrices also have
deep connections to the statistical properties of zeros of the
Riemann zeta function \cite{mehta-book}.

In the mid 1990s, supersymmetric matrix models we proposed as
non-perturbative definitions of M-theory and superstring
theory \cite{BFSS,IKKT}. Bosonic matrix models are also studied
\cite{krauth-mc-approach-m-theory,krauth-ym-integ,
ambjorn-mc-IIB,sugino-conv-gauss-exp,
cicuta-ym-integ,austing-conv-ym-integ} as a first step towards
understanding these supersymmetric matrix models.

Recently, interest in matrix models has been revived in several contexts.

Work of Dijkgraaf and Vafa
\cite{dijkgraaf-vafa-1,dijkgraaf-vafa-2} has shown that the
effective superpotential of ${\cal N}=1$ supersymmetric gauge
theories with an adjoint chiral superfield can be calculated
exactly from a bosonic one matrix model in the large $N$ limit.
This correspondence also has extensions to multi-matrix models,
for example the glueball superpotential of ${\cal N} = 1^*$ gauge
theory was determined from the partition function of the
corresponding three-matrix model
\cite{dijkgraaf-vafa-2,kazakov-3-mat,kazakov-marshakov}. Bosonic
multi-matrix models also arise in the context of quiver gauge
theories \cite{chiral-rings}.

The $c = 1$ matrix model has been revived in the context of two
dimensional super-gravity coupled to $\hat c=1$ matter
\cite{chat-equal-1-mat-model}. $c=1$ matrix models are also of
current interest in studying unstable D-branes and the phenomenon
of tachyon condensation \cite{verlinde-mcgreevy,sen-tach-cond}.

We may also regard finite matrix models as zero momentum limits of
matrix field theories such as Yang-Mills theory. They provide a testing
ground for new ideas on matrix field theories.

While matrix models have been extending their influence into a wide
variety of contexts, there have been attempts to solve matrix models
exactly. For instance, the work of Brezin et.al. \cite{bipz}, Mehta
\cite{mehta-AB}, Kazakov \cite{kazakov-ABAB}, Staudacher \cite{staudacher}
and others has shed light on the partition functions and certain
special classes of correlations in one and two matrix models. The two matrix
model studied by Mehta is actually a special case of a class of chain
multi-matrix models which can all be solved exactly, leading in the
continuum limit to the solution of the $c = 1$ matrix quantum mechanics
found by Brezin et. al.

However, it has proven difficult to determine the correlations of
a generic multi-matrix model analytically. In the 't Hooft limit
the fluctuations in invariant observables becomes small and the
theory becomes classical while retaining the quantum fluctuations
in $\hbar$ and certain other non-perturbative features. Aside from
some special cases where exact solutions for the partition
function and certain correlations have been possible, multi-matrix
models, even in the large $N$ limit are not exactly solved. This
should not be surprising, since they represent complicated
classical dynamical systems. Monte-carlo integration is probably
the best available means of obtaining numerically accurate
predictions. Their drawback is that some of the mathematical
structures are not revealed. It would be useful to have a middle
ground: an approximation method that provides both qualitative and
mathematical insights and a quantitative estimate for
correlations.

In a previous paper (\cite{entropy}, see also \cite{information,
coh-mm-eq}) we derived a variational principle for large $N$
multi-matrix models. This involved many new theoretical ideas,
especially the role played by the automorphism group of the free
algebra generated by the matrices. The most important physical
principle learnt was that the cohomology of this group is a
non-commutative entropy. It was the crucial element in deriving a
{\it classical action} for large $N$ matrix models. This allowed
us to obtain variational approximations for the correlation
tensors of multi-matrix models in the large $N$ limit. Thus we
have a self contained formulation and method of approximate
solution for matrix models in the $N \to \infty$ limit. It deals
directly with correlations, rather than matrix elements. There is
no more any reference to integrations over matrices nor the
principle of unitary invariance. Rather, we have a classical
theory on the configuration space of non-commutative probability
distributions. The correlation tensors are coordinates on this
space. They are determined by a constrained maximization of a
non-commutative analogue of entropy. The action of the original
matrix model is encoded in the constraints.

To solve this extremization problem approximately, we maximize the
entropy on a conveniently chosen finite-dimensional sub-space of
the configuration space. The simplest such choice is the subspace
corresponding to the wignerian correlations.

In \cite{entropy} we compared our variational ansatz with the
exact solution of a two matrix model studied by Mehta
\cite{mehta-AB}. The variational ansatz gave a reasonably
good approximation both for strong and weak coupling (summarized in
section \ref{s-mehta-model}). This
gives us the confidence to test our ansatz for models that
have not been exactly solved. In this paper we quantitatively test the
wignerian variational ansatz with correlations measured using
Monte-Carlo integration for a bosonic two matrix model with action
$\tr [\half m^2 (A_1^2 + A_2^2) - \ov{4} [A_1,A_2]^2]$. We study a
two matrix model since it is the simplest multi-matrix model. We
pick this action since it mimics that of the zero momentum limit
of Yang-Mills theory. Moreover, this is the simplest two matrix
model that is not exactly solved and also shares the derivation
property of Yang-Mills theory
\cite{derivation-shuffle,migdal-loop}. We use the Metropolis
algorithm implemented on a personal computer to test our new
variational principle with the simplest of ansatze. We compare
measured and variational two and four point correlations (eg.
$<\Ntr A_1 A_2 A_2 A_1>$) and also the expectation values of
Wilson loop operators (eg. $<\Ntr e^{i l A_1}e^{i l A_2}e^{-i l
A_1}e^{-i l A_2} >$) in the large N limit. We find that it works
remarkably well for $m^2$  of order $1$ or more.  As $m^2 \to 0$,
the measured correlations appear to grow without bound, reflecting
the divergence in the matrix integrals that define the pure
commutator squared model. This is not captured by the wignerian
ansatz.

We now summarize the framework we use to study
large $N$ matrix models.

\section{Large $N$ Matrix Models}

We consider multi-matrix models where the dynamical variables are
a set of $M$ hermitian $N \times N$ matrices $[A_i]^a_b$. Here $i
= 1, \cdots, M$ labels the matrices and $a,b = 1, \cdots, N$ are
the `color' row and column indices. The action $S(A)$ for the
matrix model is a $U(N)$ invariant polynomial in $S(A) = tr S^I
A_I$. \footnote{Capital letters denote multi-indices $S^I = S^{i_1
\cdots i_p},~ A_I = A_{i_1} A_{i_2} \cdots A_{i_p}$} An example is
the $2$ matrix model

\beq
     S = \tr \bigg[- {1 \over 4} [A_1, A_2]^2 \bigg]
\eeq

\fl Let

\beq
    Z =  \int dA e^{-N S(A)}
\eeq

\fl denote the partition function. The observables we are
interested in are the correlation tensors $G_I$ of the large $N$
limit, the limit where $N \to \infty$ holding the coupling
constants $S^I$ fixed.

\beqs
     <\Phi_{i_1 \cdots i_p}> &\equiv& <{\tr \over N} A_{i_1} \cdots A_{i_p}> = {1 \over Z} \int dA
    e^{-N S(A)} {\tr \over N} A_{i_1} \cdots A_{i_p} \cr
     G_{i_1 \cdots i_p} &=& \lim_{N \to \infty} <\Phi_{i_1 \cdots i_p}>
\eeqs

\fl These are a complete set of observables in the $N \to \infty$
limit since expectation values of products of invariants factorize

\beq
    <\Phi_{I_1} \Phi_{I_2} \cdots \Phi_{I_r}> =
    <\Phi_{I_1}><\Phi_{I_2}> \cdots <\Phi_{I_r}> +
    {\cal O}(\ov{N^2})
\eeq

\fl The $G_I$ satisfy factorized loop equations (factorized
Schwinger Dyson equations):

\beq
     S^{J_1 i J_2} G_{J_1 I J_2} =
     \delta_I^{I_1 i I_2} G_{I_1}G_{I_2}
\eeq

\fl On the left side is an action dependent term while on the
right is an anomalous universal term related to the
non-commutative entropy explained in \cite{entropy}.

It has been shown \cite{austing-conv-ym-integ} that the integrals
over matrices for the pure commutator squared action for a two
matrix model are not convergent. To see this, consider the
partition function, and go to the basis in which $A_1$ is
diagonal. In this basis, the integrand is independent of the
diagonal elements of $A_2$, and therefore diverges. The divergence
is even worse if we consider the expectation of the trace of a
polynomial involving $A_2$, since then the integrand grows for
large values of the diagonal elements of $A_2$. It is thus
necessary for us to regularize the action. The simplest
possibility is to add a quadratic term which ensures convergence
of the matrix integrals:

\beq
     S = \half(A_1^2 + A_2^2) - {g^2 \over 4} [A_1,A_2]^2
\eeq

\fl This is the model we focus on. There is another reason to
consider this two matrix model. An important property of
Yang-Mills theory (in any dimension) is that its action leads to
factorized loop equations whose action-dependent term is a
derivation of the shuffle product of correlation tensors
\footnote{This remark will be explained in detail in a forthcoming
paper on the algebraic structure of factorized loop equations
\cite{derivation-shuffle}.}. Therefore we would like to study a
two matrix model whose action shares this property. It can be
shown that the most general quartic \footnote{There are polynomial
interactions of higher order with the derivation property.}
M-matrix model with this property is

\beqs
    S &=& \tr \bigg[S^{i_1 i_2} A_{i_1 i_2} + S^{i_1 i_2 i_3}
    [A_{i_1},A_{i_2}] A_{i_3} + S^{i_1 i_2 i_3 i_4}(A_{i_1 i_2 i_3 i_4}
    - A_{i_2 i_1 i_3 i_4} \cr && ~~~ +~A_{i_3 i_2 i_1 i_4} - A_{i_3 i_1 i_2 i_4}) \bigg]
\eeqs

\fl where $S^I$ are arbitrary cyclically symmetric tensors. We can
use a $GL_M(\bf{C})$ change of basis $\tilde A_i = T^j_i A_j$ to
reduce the action to the canonical form where the covariance
$S^{ij} \mapsto \half \delta^i_j$. Under such a change of basis
$S(A) = S^{i_1 \cdots i_n} A_{i_1 \cdots i_n} \mapsto \tilde S(A)
= S^{j_1 \cdots j_n} T^{i_1}_{j_1} \cdots T^{i_n}_{j_n} A_{i_1
\cdots i_n}$ and $G_{i_1 \cdots i_n} = T^{j_1}_{i_1} \cdots
T^{j_n}_{i_n} \tilde G_{j_1 \cdots j_n}$. Thus the correlations
$G_I$ can be obtained from those of the canonical action. In the
two matrix case ($M=2$) the action reduces to

\beq
    S = \tr \bigg[\half (A_1^2 + A_2^2) - {g^2 \over 4} [A_1,A_2]^2 \bigg]
\eeq

\fl There is only one independent coupling constant, the ratio of
the coefficients of the quadratic and quartic terms. It is more
convenient to study

\beq
    S = \tr \bigg[{m^2 \over 2} (A_1^2 + A_2^2) - {1 \over 4} [A_1,A_2]^2 \bigg]
\eeq

\fl since it allows us to consider the pure commutator squared
model in the $m^2 \to 0$ limit. In the large $N$ limit, all
correlations are functions of the single coupling constant $m^2$.
In the pure commutator squared model, the coupling constant is an
overall factor in the action and can be scaled out, the dependence
of correlations on it can be determined by dimensional analysis.
By contrast, in the model we study, the coupling constant
dependence of the correlations is to be dynamically determined,
making it more analogous to Yang-Mills theory than the large $N$
reduced models of M-theory
\cite{krauth-mc-approach-m-theory,krauth-ym-integ,
ambjorn-mc-IIB,sugino-conv-gauss-exp,
cicuta-ym-integ,austing-conv-ym-integ}

Let us now summarize how we measure correlations by Monte-Carlo
simulation.

\section{Monte Carlo Measurement of Correlations}

To measure the correlations numerically, we generate an ensemble
of matrix configurations $A_i^{(k)}, k =1, \cdots , n$ such that
as $n \to \infty$, matrix elements $[A_i]^a_b$ picked at random
from this ensemble are distributed according to $\ov{Z} e^{-N
S(A)}$.

The Metropolis algorithm
\cite{metropolis,creutz,bhanot-metropolis,itzykson-drouffe} is
used to create such an ensemble. We begin with a configuration
$A_i^{(1)}$ such as the zero matrices. The matrix elements are
updated sequentially preserving hermiticity. At the $k^{th}$ time
step, a candidate configuration is generated $B_i = A_i^{(k)} +
W_i + W_i^\dag$. Here for each $i$, $W_i$ is a random $N \times N$
Weyl matrix. The only non-vanishing matrix element of $W_i$ is
picked at random from a uniform distribution of complex numbers in
a square whose diagonally opposite vertices are at $-\Lambda
-\sqrt{-1} \Lambda, \Lambda + \sqrt{-1} \Lambda$.

If $B$ has a greater Boltzmann weight than $A^{(k)}$, the
candidate is accepted,

\beq
  A_i^{(k+1)} = B_i {\rm ~~if~~} e^{-N(S(B)-S(A^{(k)}))} \geq 1
\eeq

\fl and when $B$ has a lesser Boltzmann weight than $A^{(k)}$, the
change is accepted with probability $e^{-N(S(B)-S(A^{(k)}))}$ and
rejected otherwise

\beqs
    A_i^{(k+1)} &=& B_i {\rm ~~if~~} 1 \geq e^{-N(S(B)-S(A^{(k)}))} > r \cr
    A_i^{(k+1)} &=& A_i^{(k)} {\rm ~~otherwise}
\eeqs

\fl Here $r$ is a random number uniformly distributed in $(0,1)$.

We can regard each matrix element as a continuous spin variable.
One Metropolis sweep corresponds to $\half (N^2 - N) + N$ time
steps of sequentially updating each independent matrix element of
the $M$ matrices. We perform a large number $n_s$ of such
Metropolis sweeps, generating an ensemble with  $n = \half n_s
(N^2 + N)$ configurations ($n = 18000$ for the measurements
presented in this paper).

It is shown \cite{creutz,bhanot-metropolis} that as $n \to \infty$
this algorithm produces a Boltzmann ensemble of configurations.
The idea is that the above rules for making a transition can be
used to define a Markov matrix on the space of ensembles. Its
eigenvalue of maximum modulus ($=1$) corresponds to the
eigenvector labelling the Boltzmann ensemble. Thus one defines a
contraction mapping on the space of ensembles whose unique fixed
point is the Boltzmann ensemble.

Once we have this ensemble, the correlations are computed:

\beq
    <{\tr \over N} A_{i_1} \cdots A_{i_p}> = \ov{n} \sum_{k=1}^n {\tr \over N}
    A_{i_1}^{(k)} \cdots A_{i_p}^{(k)}
\eeq

\fl To extract the large $N$ limit of the correlations, we do the
measurement for several values of $N = 10,7,...,15$ and fit the
results to the known $N$ dependence for large $N$:

\beq
     <{\tr \over N} A_{i_1} \cdots A_{i_p}> \rightarrow G_{i_1 \cdots
    i_p} + {\tilde G_{i_1 \cdots i_p} \over N^2}
\eeq

\fl and extract the value of $G_{i_1 \cdots i_p}$.

We mention the main sources of error in these measurements. First,
there is the statistical error (${\cal O}({variance \over
\sqrt{n}})$) of truncating the ensemble of configurations at a
finite value of $n$. This is estimated by the bootstrap
\cite{bootstrap} procedure. Next there is the systematic error
that could arise from the choice of initial configuration of
matrices. This is estimated by changing the initial configuration
slightly. The truncation ($\Lambda$) of the region in the complex
plane from where the increments to random matrix elements are
picked is a third source of error. As a practical matter we pick
$\Lambda$ so that the acceptance of the algorithm to candidate
configurations is roughly $50\%$. For sufficiently large $n$, we
find the measurements to be insensitive to small changes in the
value of $\Lambda$. The value of $\Lambda$ needs to be increased
in order to measure correlations of very high order accurately
while holding other parameters fixed. Finally, there is the error
in extracting the large $N$ limit of the correlations from finite
$N$ data.

\section{Variational Principle for Large $N$ Matrix Models}

Let us summarize the variational principle introduced in
\cite{entropy}. Given an action $S(A) = \tr S^I A_I$, for an $M$
matrix model we want to determine the correlations $G_J$ in the
large $N$ limit. We found a variational principle $\Omega(G) =
\chi(G) - S^I G_I$ whose extremization leads to the factorized
loop equations $S^{J_1 i J_2} G_{J_1 I J_2} = \delta_I^{I_1 i I_2}
G_{I_1}G_{I_2} $. The variation of $S^I G_I$ gives the action
dependent term on the left while the variation of $\chi$ gives the
universal term on the right. The main difficulty was that the
non-commutative entropy $\chi$ is not a power series in the $G_I$.
The space of correlations is a coset space of the automorphism
group of the free algebra by the subgroup of measure preserving
automorphisms. So we expressed $\chi$ as an invariant power series
on the larger space of automorphisms of the free algebra. $\chi$
is actually a non-trivial $1$-cocycle of this group. The formula
for $\chi$ is given in \cite{entropy}, we will only use a special
case of it in this paper.

Thus, to determine the correlations $G_J$, we must maximize the
entropy $\chi$ while holding the correlations $G_I$ conjugate to
the coupling tensors $S^I$ in the given action fixed. In other
words, in the factorized loop equations, $S^I$ are the Lagrange
multipliers enforcing these constraints.

The resulting maximum value $\chi_{max}$ has a simple physical
meaning. Let $S_0(A) = \half \tr \sum_{i=1}^M A_i A_i$ be the
canonical gaussian action for an $M$ matrix model. Suppose $Z$ and
$Z_0$ are the partition functions of $S$ and $S_0$. Then

\beq
     \chi_{max} = \lim_{N \to \infty} \ov{N^2} \log{\bigg[{Z \over
    Z_0} \bigg]}
\eeq

\fl i.e. $-\chi_{max}$ is the free energy or vacuum energy
measured with respect to the canonical gaussian matrix model.

Maximizing $\Omega$ exactly is equivalent to solving the
factorized loop equations exactly, which has not been possible in
general. To determine the correlations approximately, we maximize
$\Omega$ on a subspace of the configuration space. The simplest
subspace is that corresponding to the correlations of the
multivariate Wigner distribution. This is the wignerian
variational ansatz. On this subspace

\beq
    \Omega[G_{ij}] = \half \log \det[G_{ij}] - S^I G_I
\eeq

\fl where $G_I$ are the correlations of the multi-variate Wigner
distribution. They are determined in terms of the two point
correlation matrix by the planar analogue of Wick's theorem:
$G_{ijkl} = G_{ij}G_{kl} + G_{il}G_{jk}, {\rm ~etc}$. Thus we
regard the matrix elements of $G_{ij}$ as variational parameters
and maximize $\Omega$ for the given action $S$. To summarize, the
wignerian variational ansatz gives the wignerian correlations that
best approximate the true correlations of a given matrix model.
Best approximation here means the one that maximizes entropy while
holding the correlations conjugate to $S^I$ fixed.

We now compare our wignerian variational ansatz, first with the
exact solution of a two matrix model studied by Mehta and then with
the numerical solution of the gaussian + Yang-Mills two matrix model.

\section{Two Matrix Model studied by Mehta}
\label{s-mehta-model}

We recall here (for details see \cite{entropy}) the comparison
of our variational ansatz with Mehta's exact solution of a two
matrix model. In \cite{mehta-AB} Mehta finds the exact vacuum energy
of the two matrix model with action

\beq
    S(A,B) = \tr\bigg[\half(A^2 + B^2 - cAB -cBA) +
    {g \over 4}(A^4 + B^4)  \bigg]
\eeq

\fl From his solution, we may extract the exact values of

\beq
     E^{ex}(g,c) = - \lim \ov{N^2} \log{Z(g,c) \over Z(0,0)},
     ~~G^{ex}_{AB} ~~ {\rm and} ~~ G^{ex}_{AAAA}.
\eeq

\fl in the strong and weak coupling regimes. These are
compared with our variational estimates below. For small $g$ and
$c = \half$:

\beqs
    E^{ex}(g,\half) &=& -.144 + 1.78 g - 8.74 g^2 + \cdots \cr
    E^{var}(g,\half) &=& -.144 + 3.56 g - 23.7 g^2 + \cdots
    \cr \cr
    G_{AB}^{ex}(g,\half) &=& {2 \over 3} - 4.74 g + 53.33 g^2 + \cdots \cr
    G_{AB}^{var}(g,\half) &=& {2 \over 3} - 4.74 g + 48.46 g^2 + \cdots \cr
    \cr
    G_{AAAA}^{ex}(g,\half) &=& {32 \over 9} - 34.96 g + \cdots
    \cr
    G_{AAAA}^{var}(g,\half) &=& {32 \over 9} - 31.61 g + 368.02 g^2 +
    \cdots
\eeqs

\fl For strong coupling and arbitrary $c$:

\beqs
    E^{ex}(g,c) &=& \half \log g + \half \log{3}-{3 \over 4}
                + \cdots \cr
    E^{var}(g,c) &=& \half \log{g} + \half \log{2} + {1 \over \sqrt{8g}}
            + {\cal O}({1 \over g}) \cr
    \cr
    G_{AB}^{ex}(g,c) &\to& 0 {\rm ~as~} g \to \infty \cr
    G_{AB}^{var}(g,c) &=& {c \over 2g} - {c \over (2g)^{3 \over 2}} +
    {\cal O}({1 \over g^2}) \cr
    \cr
    G_{AAAA}^{ex}(g,c) &=& {1 \over g} + \cdots.
    \cr
    G_{AAAA}^{var}(g,c) &=& {1 \over g} - {2 \over (2g)^{3 \over 2}} + {\cal O}({1 \over
    g^2})
\eeqs

We see that both for strong and weak coupling, our wignerian variational
ansatz provides good estimates for the partition function and correlations.
We now consider the gaussian $+$ Yang Mills two matrix model, where we
compare our variational estimates with Monte-Calro measurements.

\section{Gaussian $+$ Yang-Mills two matrix model}

Let us specialize to the two matrix model with action ($m^2 > 0$)

\beq
    S(A) = \tr \bigg[ {m^2 \over 2} (A_1^2 + A_2^2) - {1 \over 4} [A_1,
    A_2]^2 \bigg]
\eeq

\fl For the wignerian ansatz,

\beq
    \Omega[G] = \half \log \det[G_{ij}] - {m^2 \over 2} (G_{11} + G_{22})
    + {1 \over 2} (G_{1212} - G_{1221})
\eeq

\fl Due to the $A_1 \leftrightarrow A_2$ symmetry of the action we
can assume a variational matrix

\beq
    G_{ij} = \left(%
\begin{array}{cc}
  \alpha & \beta \\
  \beta & \alpha \\
\end{array}%
\right)
\eeq

\fl Since $<\Ntr A_1^2>~ \geq 0$ and $<\Ntr (A_1-A_2)^2> ~\geq 0$,
we must maximize

\beq
    \Omega(\alpha,\beta) = \half \log(\alpha^2 - \beta^2) - m^2 \alpha
     + {1 \over 2} (\beta^2 - \alpha^2)
\eeq

\fl in the region $\alpha \geq 0$ and $\alpha \geq \beta$. We get

\beq
    G_{11} = G_{22} = \alpha = \sqrt{1+ {m^4 \over 4}} - {m^2 \over 2},~~
         G_{12} = G_{21} = \beta = 0 \label{def-of-alpha}
\eeq

\begin{figure}
\centerline{\epsfxsize=6.truecm\epsfbox{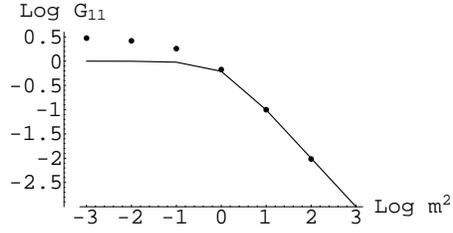}}
\caption{$\log_{10}{[G_{11}]}$ versus $\log_{10}{[m^2]}$. Solid
line is variational estimate, dots are the Monte-Carlo
measurements. The approximation becomes poor for small values of
$m^2$} \label{G11}
\end{figure}

\begin{figure}
\centerline{\epsfxsize=6.truecm\epsfbox{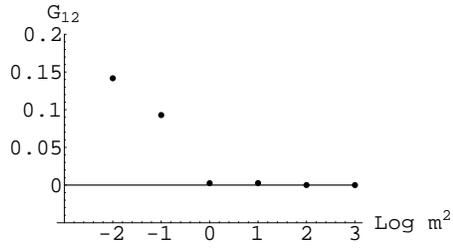}}
\caption{$ G_{12}$ versus $\log_{10}{[m^2]}$} \label{G12}
\end{figure}

\fl Figures \ref{G11} and \ref{G12} compare the variational two
point correlations with Monte-Carlo measurements for a range of
values of $m^2$.

All other correlations can be expressed in terms of these. For
example, the 4-point correlations are (the rest are determined by
cyclic symmetry and $A_1 \leftrightarrow A_2$ exchange symmetry)

\beq
    G_{1111} = 2 \alpha^2; ~~
    G_{1212} = 0; ~~
    G_{1221} = \alpha^2; ~~
    G_{1112} = 0
\eeq

\begin{figure}
\centerline{\epsfxsize=6.truecm\epsfbox{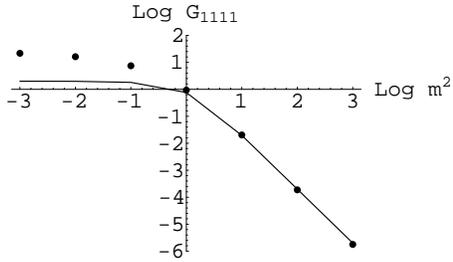}}
\caption{$ \log_{10}{[G_{1111}]}$ versus $\log_{10}{[m^2]}$}
\label{G1111}
\end{figure}

\begin{figure}
\centerline{\epsfxsize=6.truecm\epsfbox{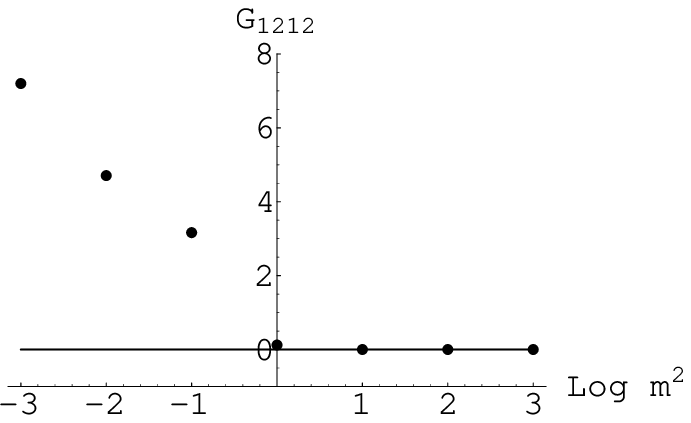}}
\caption{$ G_{1212}$ versus $\log_{10}{[m^2]}$} \label{G1212}
\end{figure}

\begin{figure}
\centerline{\epsfxsize=6.truecm\epsfbox{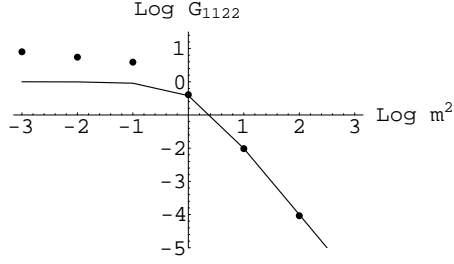}}
\caption{$ \log_{10}{[G_{1122}]}$ versus $\log_{10}{[m^2]}$}
\label{G1122}
\end{figure}

\fl Figures \ref{G1111}, \ref{G1212} and \ref{G1122} compare
variational estimates (solid lines) and Monte-Carlo measurements
(dots) of $G_{1111}$, $G_{1212}$ and $G_{1122}$ for $10^{-3} \leq
m^2 \leq 10^3$. The $n$ point pure $A_1$ (or $A_2$) correlation is
given by the Catalan numbers

\beq \label{catalan}
    G_{111 \cdots 1} = G_{222 \cdots 2} \equiv G_{(n)} = \left\{%
\begin{array}{ll}
    c_{n \over 2} \alpha^{n \over 2} = {n! \over ({n \over 2})!
        ({n \over 2}+1)!} \alpha^{n \over 2},
        & \hbox{if n is even;} \\
    0, & \hbox{if n is odd.} \\
\end{array}%
\right. \eeq

\fl More generally,

\beqs \label{useful-correlations}
    G_{11 \cdots 1 22 \cdots 2} &\equiv& G_{(n_1)(n_2)} = G_{(n_1)}
        G_{(n_2)} \cr
    G_{11 \cdots 1 22 \cdots 2 11 \cdots 1 22 \cdots 2} &\equiv&
        G_{(n_1)(n_2)(n_3)(n_4)} = G_{(n_1+n_3)} G_{(n_2)} G_{(n_4)}
        \cr &&
        + G_{(n_1)} G_{(n_3)} G_{(n_2+n_4)} - G_{(n_1)} G_{(n_2)}
        G_{(n_3)}G_{(n_4)}
\eeqs

\fl We mention these since they are useful in estimating
expectation values of Wilson loop-like operators. The variational
estimate for vacuum energy is

\beq
     E_{var}(g) = -\lim_{N \to \infty} \ov{N^2} \log{\bigg[{Z(g)
        \over Z(0)}\bigg]} = -\log \alpha = -\log{ \bigg[\sqrt{1 +
        {m^4 \over 4}} - {m^2 \over 2} \bigg]}
\eeq

\subsection{Wilson Loop Operators}

\fl It is also interesting to see what the wignerian ansatz says
about the 2-matrix analogue of the expectation of the Wilson loop
in the large $N$ limit.


\fl {\bf Wilson Line:} The simplest analogue is a `Wilson line'
the analogue of the parallel transport along a line of length $l$
in the $A_1$ direction

\beq
    W_{line}(l) \equiv \lim_{N \to \infty} <\Ntr e^{i l A_1}>
\eeq

\fl For the wignerian ansatz, we get (using eqn (\ref{catalan}),
$J_n(z)$ is the Bessel function of the first kind)

\beqs
    W_{line}(l) &=& \sum_{n=0}^\infty {(i l)^{2k} \over (2k)!} c_{k}
        \alpha^k \cr
    &=& \ov{l \sqrt{\alpha}} J_1(2 l \sqrt{\alpha}) \sim
        \ov{\sqrt{\pi} (l \sqrt{\alpha})^{3 \over 2}} \cos{({3\pi \over 4}
        - 2 l \sqrt{\alpha})} {\rm ~as~} l \to \infty
\eeqs

\begin{figure}
\centerline{\epsfxsize=6.truecm\epsfbox{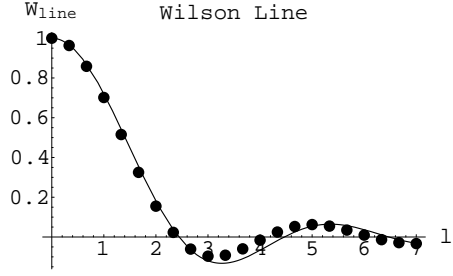}}
\caption{$W_{line}(l)$ for $m^2 = 1$. Dots are numerical and solid
line variational estimate.} \label{W-line}
\end{figure}

\fl $W_{line}(l)$ is a real-valued function of real $l$ since the
odd order correlations vanish. Thus, for the wignerian ansatz, the
expectation value of the `Wilson line' is oscillatory but decays
as a power $l^{-3/2}$. For small $l$, $W_{line}(l) \to 1 - \half
\alpha l^2 + {\alpha^2 l^4 \over 12} - \cdots$. Figure
\ref{W-line} compares this ansatz with Monte-Carlo measurements
for $m^2 = 1$. The behavior both for small and large values of $l$
is well captured by our ansatz.

\fl {\bf L shaped Wilson Line:} For an $L$ shaped curve, we define

\beq
    W_{L}(l)=\lim_{N \to \infty} <\Ntr e^{ilA_1} e^{ilA_2}>
\eeq

\fl For the wignerian ansatz (use eq. (\ref{useful-correlations});
$_1F_2(a,{\bf b};z) = \sum_{n=0}^\infty {(a)_n \over (b_1)_n
(b_2)_n} {z^n \over n!}$ is a generalized Hypergeometric function
with $(a)_n$ the Pochhammer symbol),

\beqs
     W_{L}(l) &=& \sum_{n_1, n_2 =0}^\infty {(il)^{n_1 +n_2}
        \over n_1! n_2!} \lim_{N \to \infty} <\Ntr A_1^{n_1} A_2^{n_2}> \cr
     &=& \sum_{k_1,k_2 =0}^\infty {(-l^2 \alpha)^{k_1 + k_2} \over
        k_1! (k_1 +1)! k_2! (k_2 +1)!} \cr
     &=& \sum_{n=0}^\infty (-l^2 \alpha)^n {4^{n+1} \Gamma(n+ {3 \over 2})
        \over \sqrt{\pi} ~\Gamma(n+1) \Gamma(n+2) \Gamma(n+3)} \cr
     W_L(l) &=& _1F_2({3 \over 2}; \{2,3 \}; -4 l^2 \alpha)
\eeqs

\begin{figure}
\centerline{\epsfxsize=6.truecm\epsfbox{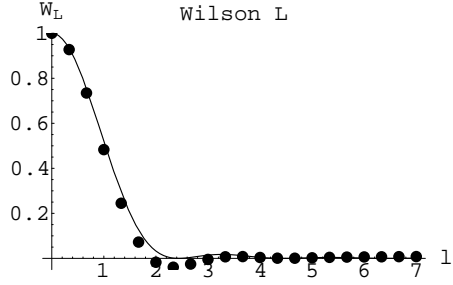}}
\caption{$W_{L}(l)$ for $m^2 =1$} \label{W-L}
\end{figure}

\fl As before $W_{L}(l)$ is real for real $l$. For small $l$,
$W_L(l) \to 1 - \alpha l^2 + {5 \alpha^2 l^4 \over 12} - \cdots$.
This is compared with the numerical calculation in fig. \ref{W-L}
for $m^2 = 1$. Both the small $l$ behavior and decay for large $l$
are captured by our estimate.

\fl {\bf Wilson Square:} The analogue of the parallel transport
around a square of side $l$ in the $A_1 - A_2$ plane is

\beq
    W_{square}(l) = \lim_{N \to \infty} <\Ntr e^{i l A_1} e^{i l A_2}
        e^{-i l A_1} e^{-i l A_2}>
\eeq

\fl In the wignerian variational approximation, $W_{square}(l)$ is
real-valued since odd order correlations vanish. Using
eq.(\ref{useful-correlations}) we get

\beq
    W_{square}(l) = \sum_{n=0}^\infty (-l^2 \alpha)^n T_1(n) + 2
            (l^2 \alpha) \sum_{n=0}^\infty (-l^2 \alpha)^n T_2(n)
\eeq

\fl where,

\beqs
    T_1(n) &=& \sum_{k_i \geq 0, k_1+\cdots +k_4=n} {2 c_{k_1 + k_3} c_{k_2} c_{k_4}
        - \Pi_{i=1}^4 c_{k_i} \over \Pi_{i=1}^4 (2k_i)!} \cr
    T_2(n) &=& \sum_{k_i \geq 0, k_1+\cdots +k_4=n} {c_{k_1 +k_3 +1} c_{k_2} c_{k_4}
        \over (2k_1 +1)! (2k_2)! (2k_3 +1)! (2k_4)!}
\eeqs

\begin{figure}
\centerline{\epsfxsize=6.truecm\epsfbox{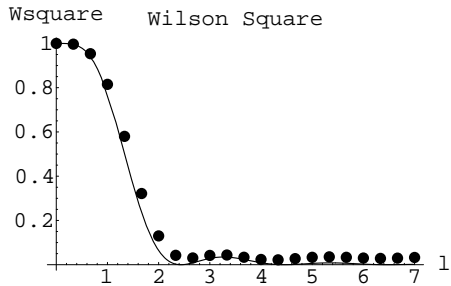}}
\caption{$W_{square}(l)$ for $m^2 = 1$} \label{W-square}
\end{figure}

\fl For small $l$, $W_{square}(l) \to 1 - l^4 \alpha^2 + {5 l^6
\alpha^3 \over 6} - \cdots$. The expectation value of the Wilson
loop is a rapidly decaying function for large values of $l$. It
would be interesting to find the asymptotic rate of decay. It is
oscillatory but a positive function, unlike $W_{line}$. These
variational predictions are confirmed by the numerics, in fig.
\ref{W-square} for $m^2 = 1$.

The variational ansatz does a very good job of estimating the
Wilson loop averages over the entire range of values of $l$
studied, for $m^2$ of order unity or more.

\section{Summary and Discussion}

We find that despite its simplicity, the
wignerian ansatz for our entropic variational principle is
remarkably good at estimating correlations. For the
exactly solved model studied by Mehta, our ansatz works well
both for strong and weak coupling. For the gaussian + Yang-Mills
two matrix model, our estimates for correlations and expectation
values of Wilson loop operators are accurate for moderate and weak
coupling when compared with Monte-Carlo measurements. When
the commutator squared term dominates the action,
the wignerian ansatz becomes poorer as an
approximation. This is to be expected. The measured correlations
grow without bound as we approach the pure commutator squared limit.
This reflects the divergence of
the matrix integrals for the pure commutator squared interaction.

We can go beyond the wignerian ansatz for the entropic variational
principle. In particular, this will allow us to improve on the
wignerian ansatz and also estimate correlations that are
identically zero for the wignerian ansatz. We will address this
question in a future paper.

In another direction, we hope to extend our approximation methods
to supersymmetric matrix models in order to make predictions about
the matrix models of M-theory and superstring theory mentioned in
the introduction.

A conceptual shortcoming of the numerical procedure used in this
paper is that we measured correlations for several finite values
of $N$ before extrapolating to the $N=\infty$ limit. Is there some
way of computationally determining the $N=\infty$ correlations
directly? This is a challenging problem that would likely require
new ideas from non-commutative algebra, geometry, probability
theory and computer science to determine directly the $N = \infty$
correlation tensors without integrating over matrix elements.

It would be interesting to make precise the field theoretic
connection between general multi-matrix models and supersymmetric
gauge theories with adjoint chiral superfields. Our methods for
estimating the correlations of the former can potentially shed
light on the expectation values of operators in the chiral ring of
the supersymmetric gauge theory.

Our investigations have focussed on matrix models with a finite
number of matrices. It would be desirable to have a similar
theoretical approach based on approximation methods over and above
the large $N$ limit, for matrix field theories (beyond $c=1$
matrix quantum mechanics) such as Yang-Mills theory. This would
complement the lattice gauge theory Monte-Carlo efforts (see eg.
\cite{teper-lattice-qcd}).

Acknowledgements: The author thanks S.G. Rajeev, A. Agarwal and L.
Akant for useful discussions and A. Agarwal for useful comments on
the manuscript.


\end{document}